\documentclass[10pt,A4paper,conference]{IEEEtran}
\usepackage{amsmath}
\usepackage{amssymb}
\usepackage{upref}
\usepackage{amsfonts}
\usepackage{graphicx}
\usepackage{verbatim}

\newcommand{\field}[1]{\mathbb{#1}}

\newcommand{\Z}{\field{Z}}

\newcommand{\R}{\field{R}}

\newcommand{\cA}{{\cal A}}

\newcommand{\cG}{{\cal G}}

\newcommand{\cP}{{\cal P}}

\newcommand{\cS}{{\cal S}}
\newcommand{\cR}{{\cal R}}
\newcommand{\cU}{{\cal U}}
\newcommand{\cT}{{\cal T}}

\newcommand{\sP}{\cP}
\newcommand{\sG}{\cG}

\newcommand{\Gr}{\smash{{\sG\kern-1.5pt}_q\kern-0.5pt(n,k)}}
\newcommand{\Grtwo}{\smash{{\sG\kern-1.5pt}_2\kern-0.5pt(n,k)}}
\newcommand{\Gkone}{\smash{{\sG\kern-1.5pt}_q\kern-0.5pt(n,k_1)}}
\newcommand{\Gktwo}{\smash{{\sG\kern-1.5pt}_q\kern-0.5pt(n,k_2)}}
\newcommand{\Ps}{\smash{{\sP\kern-2.0pt}_q\kern-0.5pt(n)}}

\newcommand{\bi}{{\bf i}}

\newcommand{\deff}{\mbox{$\stackrel{\rm def}{=}$}}

\newtheorem{theorem}{Theorem}

\newtheorem{lemma}{Lemma}

\newtheorem{example}{Example}

\begin{document}

\title{Folding, Tiling, and Multidimensional Coding}

\author{\authorblockN{Tuvi Etzion}
\authorblockA{Department of Computer Science\\
Technion-Israel Institute of Technology\\
Haifa 32000, Israel \\
Email: etzion@cs.technion.ac.il}}

\maketitle
\begin{abstract}
Folding a sequence $S$ into a multidimensional box is a method
that is used to construct multidimensional codes. The well known
operation of folding is generalized in a way that the sequence $S$
can be folded into various shapes. The new definition of folding
is based on lattice tiling and a direction in the $D$-dimensional
grid. There are potentially $\frac{3^D-1}{2}$ different folding
operations. Necessary and sufficient conditions that a lattice
combined with a direction define a folding are given. The
immediate and most impressive application is some new lower bounds
on the number of dots in two-dimensional synchronization patterns.
This can be also generalized for multidimensional synchronization
patterns. We show how folding can be used to construct
multidimensional error-correcting codes and to generate
multidimensional pseudo-random arrays.
\end{abstract}

\section{Introduction}
\label{sec:introduction}

Multidimensional coding in general and two-dimensional coding in
particular is a subject which attract lot of attention in the last
three decades. It includes error-correcting
codes~\cite{AMT,BBZS,ScEt05,EtYa09}, synchronization
patterns~\cite{GoTa82,GoTa84,Rob85,Rob97,BEMP08}, perfect maps and
pseudo-random arrays~\cite{McSl76,Etz88}, and other topics as
well. But, although the related theory of the one-dimensional case
is well developed, the theory for the multidimensional case is
developed rather slowly. This is due that the fact the most of the
one-dimensional techniques are not generalized easily to higher
dimensions. Hence, specific techniques have to be developed for
multidimensional coding. One technique that was used for
multidimensional coding is folding~\cite{Rob97,BEMP08,McSl76}. It
was used to form a two-dimensional array, in the shape of a
rectangle, from a one-dimensional sequence. It is generalized for
multidimensional arrays to form a multidimensional box.

In this paper we generalize the definition of folding. It is
generalized in a way that we will be able to apply it to a
multidimensional shape $\cS$ which does not necessarily has a
shape of a box. We present applications of the new definition to
form multidimensional synchronization patterns, error-correcting
codes, and pseudo-random arrays, for many types of
multidimensional shapes.

The rest of this paper is organized as follows. In
Section~\ref{sec:FoldTile} we define the basic concepts of folding
and lattice tiling. Tilings and lattices are basic combinatorial
and algebraic structures. We will consider only integer lattice
tilings. We will summarize the important properties of lattices
and lattice tilings. In Section~\ref{sec:folding} we will present
the generalization of folding into multidimensional shapes. We
will show that all previous known foldings are special cases of
the new definition. The new definition involves a lattice tiling
and a direction. We will prove necessary and sufficient conditions
that a lattice with a direction define a folding. In
Section~\ref{sec:bounds} we give a short summary on
synchronization patterns and present basic theorems concerning the
bounds on the number of elements in such patterns. In
Section~\ref{sec:DDAs} we apply the results of the previous
sections to obtain new type of synchronization patterns which are
asymptotically either optimal or almost optimal. In
Section~\ref{sec:ECC} we show how folding can be applied to
construct multidimensional error-correcting codes. In
section~\ref{sec:pseudo-random} we generalize the construction
in~\cite{McSl76} to form pseudo-random arrays on different
multidimensional shapes. We conclude in
Section~\ref{sec:conclude}.

\section{Folding and Lattice Tiling}
\label{sec:FoldTile}

\subsection{Folding}

Folding a rope, a ruler, or any other feasible object is a common
action in every day life. Folding an one-dimensional sequence into
a $D$-dimensional array is very similar, but there are a few
variants. First, we will summarize three variants for folding of
an one-dimensional sequence $s_0 s_1 \cdots s_{m-1}$ into a
two-dimensional array $\cA$. The generalization for a
$D$-dimensional array is straightforward while the description
becomes more clumsy.

\begin{enumerate}
\item[\bf F1.] $\cA$ is considered as a cyclic array horizontally
and vertically in such a way that a walk diagonally visits all the
entries of the array. The elements of the sequence are written
along the diagonal of the $r \times t$ array $\cA$. This folding
works if and only if $r$ and $t$ are relatively primes.

\item[\bf F2.] The elements of the sequence are written row by row
(or column by column) in $\cA$.

\item[\bf F3.] The elements of the sequence are written diagonal
by diagonal in $\cA$.
\end{enumerate}

F1 and F2 were used by MacWilliams and Sloane~\cite{McSl76} to
form a pseudo-random arrays. F2 was used by Robinson~\cite{Rob97}
to fold a one-dimensional ruler into a two-dimensional Golomb
rectangle. The generalization to higher dimensions is straight
forward. F3 was used in~\cite{BEMP08} to obtain a synchronization
patterns in the square grid.

\subsection{Tiling}

Tiling is one of the most basic concepts in combinatorics. We say
that a $D$-dimensional shape $\cS$ tiles the $D$-dimensional space
$\Z^D$ if disjoint copies of $\cS$ cover $\Z^D$. This cover of
$\Z^D$ with disjoint copies of $\cS$ is called {\it tiling} of
$\Z^D$ with $\cS$. For each shape $\cS$ we distinguish one of the
points of $\cS$ to be the {\it center} of $\cS$. Each copy of
$\cS$ in a tiling has the center in the same related point. The
set $\cT$ of centers in a tiling defines the tiling, and hence the
tiling is denoted by the pair $(\cT,\cS)$. Given a tiling
$(\cT,\cS)$ and a grid point $(i_1,i_2,\ldots,i_D)$ we denote by
$c(i_1,i_2,\ldots,i_D)$ the center of the copy of $\cS$ for which
$(i_1,i_2,\ldots,i_D) \in \cS$. We will also assume that the
origin is a center of a copy of $\cS$.
\begin{lemma}
\label{lem:center} For a given tiling $(\cT,\cS)$ and a point
$(i_1,i_2,\ldots,i_D)$ the point
$(i_1,i_2,\ldots,i_D)-c(i_1,i_2,\ldots,i_D)$ belongs to the shape
$\cS$ whose center is in the origin.
\end{lemma}

One of the most common types of tiling is a {\it lattice tiling}.
A {\it lattice} $\Lambda$ is a discrete, additive subgroup of the
real $D$-space $\R^D$. W.l.o.g., we can assume that

\begin{equation}
\label{eq:lattice_def} \Lambda = \{ u_1 v_1 + u_2v_2 + \cdots +
u_D v_D ~:~ u_1, \ldots,u_D \in \Z \}
\end{equation}
where $\{ v_1,v_2,\ldots,v_D \}$ is a set of linearly independent
vectors in $\R^D$. A lattice $\Lambda$ defined by
(\ref{eq:lattice_def}) is a sublattice of $\Z^D$ if and only if
$\{ v_1,v_2,\ldots,v_D \} \subset \Z^D$. We will be interested
solely in sublattices of $\Z^D$. The vectors $v_1,v_2,\ldots,v_D$
are called {\it basis} for $\Lambda \subseteq \Z^D$, and the $D
\times D$ matrix
$$
{\bf G}=\left[\begin{array}{cccc}
v_{11} & v_{12} & \ldots & v_{1D} \\
v_{21} & v_{22} & \ldots & v_{2D} \\
\vdots & \vdots & \ddots & \vdots\\
v_{D1} & v_{D2} & \ldots & v_{DD} \end{array}\right]
$$
having these vectors as its rows is said to be the {\it generator
matrix} for $\Lambda$.

The {\it volume} of a lattice $\Lambda$, denoted $V( \Lambda )$,
is inversely proportional to the number of lattice points per unit
volume. More precisely, $V( \Lambda )$ may be defined as the
volume of the {\it fundamental parallelogram} $\Pi(\Lambda)$ in
$\R^D$, which is given by
$$
\Pi(\Lambda) \deff\ \{ \xi_1 v_1  + \xi_2 v_2 +  \cdots + \xi_D
v_D :  0 \leq \xi_i < 1, ~ ,1 \leq i \leq D \}
$$
There is a simple expression for the volume of $\Lambda$, namely,
$V(\Lambda)=| \det {\bf G} |$.

We say that $\Lambda$ induces a {\it lattice tiling} of $\cS$ if
the lattice points can be taken as the set $\cT$ to form a tiling
$(\cT,\cS)$.

\section{The Generalized Folding Method}
\label{sec:folding}

In this section we will generalize the definition of folding. All
the previous three definitions are special cases of the new
definition. The new definition involves a lattice tiling
$(\cT,\cS)$, where $\cS$ is the shape on which the folding is
performed.

A {\it ternary vector} of length $D$, $(d_1,d_2,\ldots,d_D)$, is a
word of length $D$, where $d_i \in \{ -1,0,+1 \}$.

Let $\cS$ be a $D$-dimensional shape and let
$\delta=(d_1,d_2,\ldots,d_D)$ be a nonzero ternary vector of
length $D$. Let $(\cT,\cS)$ be a lattice tiling defined by a
$D$-dimensional lattice $\Lambda$, and let $\tilde{\cS}$ be the
copy of $\cS$ in $(\cT,\cS)$ which includes the origin. We define
recursively a {\it folded-row} starting in the origin. If the
point $(i_1,i_2, \ldots ,i_D )$ is in $\tilde{\cS}$ then the next
point on its folded-row is defined as follows:
\begin{itemize}
\item If the point $(i_1+d_1,i_2+d_2,\ldots,i_D+d_D)$ is in
$\tilde{\cS}$ then it is the next point on the folded-row.

\item If the point $(i_1+d_1,i_2+d_2,\ldots,i_D+d_D)$ is in
$\tilde{\cS}' \neq \tilde{\cS}$ whose center is in the point
$(c_1,c_2,\ldots,c_D)$ then
$(i_1+d_1-c_1,i_2+d_2-c_2,\ldots,i_D+d_D-c_D)$ is the next point
on the folded-row.
\end{itemize}

The new definition of folding is based on a lattice $\Lambda$, a
shape $\cS$, and a direction $\delta$. The triple
$(\Lambda,\cS,\delta)$ defines a folding if the definition yields
a folded-row which includes all the elements of $\cS$. It appears
that only $\Lambda$ and $\delta$ determines whether the triple
$(\Lambda,\cS,\delta)$ defines a folding. The role of $\cS$ is
only in the order of the elements in the folded-row; and of course
$\Lambda$ must define a lattice tiling for $\cS$.

How many different folded-rows do we have? In other words, how
many different folding operations can be defined? There are
$3^D-1$ non-zero ternary vectors. If $\Lambda$ with the ternary
vector $(d_1,d_2,\ldots,d_D)$ define a folding then also $\Lambda$
with the vector $(-d_1,-d_2,\ldots,-d_D)$ define a folding. The
two folded-rows are in reverse order, and hence they will be
considered to be equal. Other than these pairs of folded-rows, we
don't know whether for each $D$, there exists a $D$-dimensional
shape $\cS$ with $\frac{3^D-1}{2}$ different folded-rows. An
example for $D=2$ is given next.

\begin{example}
Let $\Lambda$ be the lattice whose generator matrix given by the
matrix
$$
G=\left[\begin{array}{cc}
3 & 2 \\
7 & 1
\end{array}\right]~.
$$
One can verify that shapes tiled by this lattice have different
folded-rows. It can be proved that this is the lattice with the
smallest volume which has this property.
\end{example}

The first two lemmas are an immediate consequence of the
definitions and provide us a concise condition whether the triple
$(\Lambda,\cS,\delta)$ defines a folding.

\begin{lemma}
Let $(\cT,\cS)$ be a lattice tiling defined by the $D$-dimensional
lattice $\Lambda$ and let $\delta = (d_1,d_2,\ldots,d_D)$ be a
nonzero ternary vector. $(\Lambda,\cS,\delta)$ defines a folding
if and only if the set $\{ (i \cdot d_1,i \cdot d_2,\ldots,i \cdot
d_D)-c(i \cdot d_1,i \cdot d_2,\ldots,i \cdot d_D) ~:~ 0 \leq i <
| \cS | \}$ contains $| \cS |$ distinct elements.
\end{lemma}

\begin{lemma}
Let $(\cT,\cS)$ be a lattice tiling defined by the $D$-dimensional
lattice $\Lambda$ and let $\delta = (d_1,d_2,\ldots,d_D)$ be a
nonzero ternary vector. $(\Lambda,\cS,\delta)$ defines a folding
if and only if $(|\cS| \cdot d_1,\ldots,|\cS| \cdot d_D)-c(|\cS|
\cdot d_1,\ldots,|\cS| \cdot d_D)=(0,\ldots,0)$ and for each $i$,
$0 < i < |\cS|$ we have $(i \cdot d_1,\ldots,i \cdot d_D)-c(i
\cdot d_1,\ldots,i \cdot d_D) \neq (0, \ldots ,0)$.
\end{lemma}

Before considering the general $D$-dimensional case we want to
give a simple condition to check whether the triple
$(\Lambda,\cS,\delta)$ defines a folding in the two-dimensional
case. For each one of the four possible folding definitions we
will give a necessary and sufficient condition that the triple
$(\Lambda,\cS,\delta)$ defines a folding.

\begin{theorem}
\label{thm:cond_fold2D} Let $\Lambda$ be a lattice whose generator
matrix is given by
$$
G=\left[\begin{array}{cc}
v_{11} & v_{12} \\
v_{21} & v_{22}
\end{array}\right]~.
$$
If $\Lambda$ defines a lattice tiling for the shape $\cS$ then the
triple $(\Lambda,\cS,\delta)$ defines a folding

\begin{itemize}
\item with the ternary vector $\delta =(+1,+1)$ if and only if
$\text{g.c.d.}(v_{22}-v_{21},v_{11}-v_{12})=1$;

\item with the ternary vector $\delta =(+1,-1)$ if and only if
$\text{g.c.d.}(v_{22}+v_{21},v_{11}+v_{12})=1$;

\item with the ternary vector $\delta =(+1,0)$ if and only if
$\text{g.c.d.}(v_{12},v_{22})=1$;

\item with the ternary vector $\delta =(0,+1)$ if and only if
$\text{g.c.d.}(v_{11},v_{21})=1$.
\end{itemize}
\end{theorem}
\vspace{0.2cm}

Theorem~\ref{thm:cond_fold2D} is generalized for the
$D$-dimensional case as follows. Let $\Lambda$ be a
$D$-dimensional lattice tiling for the shape $\cS$ with the
following generator matrix.

$$
G=\left[\begin{array}{cccc}
v_{11} & v_{12} & \ldots & v_{1D} \\
v_{21} & v_{22} & \ldots & v_{2D} \\
\vdots & \vdots & \ddots & \vdots\\
v_{D1} & v_{D2} & \ldots & v_{DD} \end{array}\right]~.
$$

Assume we have the direction vector $\delta = (d_1 ,d_2, \ldots ,
d_D)$. W.l.o.g. we assume that the first $\ell_1 \geq 1$ values
are +1's, the next $\ell_2$ values are -1's, and the last
$D-\ell_1-\ell_2$ values are 0's. There exist integer coefficients
$\alpha_1,\alpha_2,\ldots,\alpha_D$ such that

\begin{eqnarray*}
\sum_{j=1}^D \alpha_j
(v_{j1},v_{j2},\ldots,v_{jD})~~~~~~~~~~~~~~~~~~~~~~~~~~\\
=(|\cS|,\ldots,|\cS|,-|\cS|,\ldots,-|\cS|,0,\ldots,0),
\end{eqnarray*}
and there is no integer $i$, $0 < i < |\cS|$, and integer
coefficients $\beta_1,\beta_2,\ldots,\beta_D$ such that

$$
\sum_{j=1}^D \beta_j
(v_{j1},v_{j2},\ldots,v_{jD})=(i,\ldots,i,-i,\ldots,-i,0,\ldots,0)
$$

Hence we have the following $D$ equations:

\begin{equation*}
\label{eq:set+1} \sum_{j=1}^D \alpha_j v_{jr} = |\cS|,~~~ 1 \leq r
\leq \ell_1 ,
\end{equation*}
\begin{equation*}
\label{eq:set-1} \sum_{j=1}^D \alpha_j v_{jr} = -|\cS|,~~~ \ell_1
+1 \leq r \leq \ell_1 + \ell_2 ,
\end{equation*}
\begin{equation*}
\label{eq:set0} \sum_{j=1}^D \alpha_j v_{jr} = 0,~~~
\ell_1+\ell_2+1 \leq r \leq D ,
\end{equation*}

which are equivalent to the following $D$ equations:
$$
\sum_{j=1}^D \alpha_j v_{j1} = |\cS|,
$$
\begin{equation*}
\label{eq:set+1a} \sum_{j=1}^D \alpha_j (v_{jr}-v_{j1}) = 0,~~~ 2
\leq r \leq \ell_1 ,
\end{equation*}
\begin{equation*}
\label{eq:set-1a} \sum_{j=1}^D \alpha_j (v_{jr}+v_{j1}) = 0,~~~
\ell_1 +1 \leq r \leq \ell_1 + \ell_2,
\end{equation*}
\begin{equation*}
\label{eq:set0a} \sum_{j=1}^D \alpha_j v_{jr} = 0,~~~
\ell_1+\ell_2+1 \leq r \leq D .
\end{equation*}

We define now a set of $D(D-1)$ new coefficients $u_{rj}$, $2 \leq
r \leq D$, $1 \leq j \leq D$, as follows:

$$
u_{rj}=v_{jr}-v_{j1}~~ \text{for}~ 2 \leq r \leq \ell_1 ,
$$
$$
u_{rj}=v_{jr}+v_{j1}~~ \text{for}~ \ell_1+1 \leq r \leq
\ell_1+\ell_2 ,
$$
$$
u_{rj}=v_{jr}~~ \text{for}~ \ell_1+\ell_2+1 \leq r \leq D .
$$

Consider the $(D-1) \times D$ matrix
$$
H=\left[\begin{array}{cccc}
u_{21} & u_{22} & \ldots & u_{2D} \\
u_{31} & u_{32} & \ldots & u_{3D} \\
\vdots & \vdots & \ddots & \vdots\\
u_{D1} & u_{D2} & \ldots & u_{DD} \end{array}\right] .
$$

Using the Cramer's rule it is easy to verify that the following
assignment is the unique solution for the $\alpha_i$'s.

$$
\alpha_i = (-1)^{i-1}\det H_i ,~~~ 1 \leq i \leq D,
$$
where $H_i$ is the $(D-1) \times (D-1)$ matrix obtained from $H$
by deleting column $i$ of $H$.

\begin{theorem}
If $\Lambda$ define a lattice tiling for the shape $\cS$ then the
triple $(\Lambda,\cS,\delta)$ defines a folding if and only if
$\text{g.c.d.}(\det H_1 , \det H_2 , \ldots , \det H_D)=1$.
\end{theorem}

One important tool that we will use to find an appropriate folding
for a shape $\cS'$ is to use a folding of a simpler shape $\cS$
with the same volume and apply iteratively the following theorem.

\begin{theorem}
\label{thm:trans_shape} Let $\Lambda$ be a lattice, $\cS$ a
$D$-dimensional shape, $\delta=(d_1,d_2,\ldots,d_D)$ a direction,
and $(\Lambda,\cS,\delta)$ defines a folding. Assume the origin is
a point in $\cS$, $(i_1,i_2,\ldots,i_D) \in \cS$,
$(i_1+d_1,i_2+d_2,\ldots,i_D+d_D) \in \tilde{\cS}$, $\cS \neq
\tilde{\cS}$,  and the center of $\tilde{\cS}$ is the point
$(c_1,c_2,\ldots,c_D)$. Then for the shape $\cS'=\cS \cup
\{(i_1+d_1,i_2+d_2,\ldots,i_D+d_D)\} \setminus
\{(i_1+d_1-c_1,i_2+d_2-c_2,\ldots,i_D+d_D-c_D)\}$ the triple
$(\Lambda,\cS',\delta)$ also defines a folding.
\end{theorem}

\section{Bounds on Synchronization Patterns}
\label{sec:bounds}

Our motivation for the generalization of the folding operation
came from the design of two dimensional synchronization patterns.
Given a grid (square or hexagonal) and a shape $\cS$ on the grid,
we would like to find what is the largest set $V$ of dots on grid
points, $|V|=m$, located in $\cS$, such that the following
property holds. All the $\binom{m}{2}$ lines between dots are
distinct either in length or in slope. Such a shape $\cS$ with
dots is called a {\it distinct difference configuration} (DDC). If
$\cS$ is an $m \times m$ array with a dot in each row and a dot in
each column than $\cS$ is called a Costas array~\cite{GoTa82}. If
$\cS$ is a $k \times m$ array with a dot in each column then $\cS$
is called a sonar sequence~\cite{GoTa82}. If $\cS$ is a $k \times
n$ array then $\cS$ is called a Golomb rectangle~\cite{Rob85}.
These patterns have various applications as described
in~\cite{GoTa82}. A new application of these patterns to the
design of key predistribution scheme for wireless sensor networks
was described lately in~\cite{BEMP}. In this application the shape
$\cS$ might be a Lee sphere, an hexagon, or a circle, and
sometimes another regular polygon. This application requires in
some cases to consider these shapes in the hexagonal grid. F3 was
used for this application in~\cite{BEMP08} to form a DDC whose
shape is a rectangle rotated in 45 degrees in the square grid.
Henceforth, we assume that our grid is the square grid, unless
stated otherwise.

Let $\cS$ and $\cS'$ be two-dimensional shapes in the grid. We
will denote by $\Delta (\cS,\cS')$ the largest intersection
between $\cS$ and $\cS'$. $|\cS|$ will denote the number of grid
points in $\cS$. Let $m$ be a given integer. An infinite set of
dots in the grid such that each given shape $\cS$ on the grid is a
DDC with $m$ dots will be called an {\it infinite $\cS$-DDC}. The
following theorems are generalization of similar theorem
in~\cite{BEMP08}.

\begin{theorem}
\label{thm:infinite} Assume we are given an infinite $\cS$-DDC
with $m$ dots on the grid. Let $\cR$ be another shape on the grid.
Then there exists a copy of $\cR$ on the grid with at least
$\frac{m}{|\cS|} \Delta (\cS,\cR)$ dots.
\end{theorem}

\begin{theorem}
\label{thm:mediator} Assume we are given an infinite $\cS$-DDC
with $m$ dots on the grid. Let $\cR$ and $\cU$ be another shapes
on the grid. Then there exists a copy of $\cU$ on the grid with at
least $\frac{m}{|\cS| \cdot |\cR|} \Delta (\cS,\cR) \cdot \Delta
(\cR,\cU)$ dots.
\end{theorem}

In order to apply Theorem~\ref{thm:infinite} and
Theorem~\ref{thm:mediator} we will use folding of the sequences,
defined as follows, in our shape $\cS$. Let $A$ be an abelian
group, and let $E=\{a_1,a_2,\ldots,a_m\}\subseteq A$ be a sequence
of $m$ distinct elements of $A$. We say that $E$ is a
\emph{$B_2$-sequence over $A$} if all the sums $a_{i_1}+a_{i_2}$
with $1\leq i_1\leq i_2\leq m$ are distinct. For a survey on
$B_2$-sequences and their generalizations the reader is referred
to~\cite{Bry04}. The following lemma is well known and can be
readily verified.

\begin{lemma}
\label{lem:B2_diff} A subset $E=\{a_1,a_2,\ldots,a_m\}\subseteq A$
is a $B_2$-sequence over $A$ if and only if all the differences
$a_{i_1}-a_{i_2}$ with $1\leq i_1 \neq i_2\leq m$ are distinct in
$A$.
\end{lemma}

Note that if $D$ is a $B_2$-sequence over $\Z_n$ and $a\in\Z_n$,
then so is the shift $a+E=\{a+e:e\in E\}$. The following theorem,
due to Bose~\cite{Bose42}, shows that large $B_2$-sequences over
$\Z_n$ exist for many values of $n$.

\begin{theorem}
\label{thm:Bose} Let $q$ be a prime power. Then there exists a
$B_2$-sequence $a_1,a_2,\ldots,a_m$ over $\Z_n$ where $n=q^2-1$
and $m=q$.
\end{theorem}

The importance of folding a $B_2$ sequence $S$ into a given shape
$\cS$ is given by the following theorem.

\begin{theorem}
\label{thm:fold_B2} Let $\Lambda$ be a lattice, $\cS$, $n=|\cS|$,
a $D$-dimensional shape, and $\delta$ a direction. Let $E$ be a
$B_2$-sequence over $\Z_n$. If $(\Lambda,\cS,\delta)$ defines a
folding then the folded-row is a $D$-dimensional DDC. Moreover,
this DDC can be extended to infinite $\cS$-DDC.
\end{theorem}

In the sequel we will use Theorem~\ref{thm:infinite},
Theorem~\ref{thm:mediator}, and Theorem~\ref{thm:fold_B2} to form
DDCs with various given shapes with a large number of dots. To
examine how good are our bounds on the number of dots we should
know what is the upper bound on the number of dots in a DDC whose
shape is $\cS$. It was shown in~\cite{BEMP08} that for a DDC whose
shape is a regular polygon or a circle, the maximal number of dots
is at most $\sqrt{s} +o(\sqrt{s})$, when the shape contains $s$
points of the grid.

\section{Bounds for Specific Shapes}
\label{sec:DDAs}

In this section we will present some lower bounds on the number of
dots in some two-dimensional DDCs with specific shapes. One of the
main keys of our constructions, and the use of the given theory,
is the ability to produce a DDC with a rectangle shape and any
given ratio between its edges.
\begin{theorem}
\label{thm:ratio_rec} For each positive number $\gamma$ there
exist two integers $a$ and $b$ such that $\frac{b}{a} \approx
\gamma$ and an infinite $\cS$-DDC with $\sqrt{a \cdot b} R + o(R)$
dots whose shape is an $\alpha \times \beta = (b R +o(R)) \times
(a R +o(R))$ rectangle, $\alpha \beta =p^2-1$ for some prime $p$,
and even $\alpha$.
\end{theorem}

Now, we can give a few examples for specific shapes. To have some
comparison, let the radius of the circle or the regular polygons
be $R$ (the {\it radius} is the distance from the center of the
regular polygon to any one its vertices).

\subsection{Regular Hexagon in the Square Grid}
\label{sec:hex}

By Theorem~\ref{thm:ratio_rec} there exists an infinite $\cS$-DCC,
where $\cS$ is an $\alpha \times \beta = (aR +o(R)) \times
(bR+o(R))$ rectangle, such that $\frac{b}{a} \approx
\frac{\sqrt{3}}{2}$, $\alpha \beta = p^2 -1$ for some prime $p$,
and even $\alpha$. Let $\Lambda$ be the a lattice tiling for $\cS$
with the generator matrix
$$
G=\left[\begin{array}{cc}
\beta & \frac{\alpha}{2}+\theta \\
0 & \alpha
\end{array}\right]~,
$$
where $\theta=1$ if $\alpha \equiv 0~(mod~4)$ and $\theta=2$ if
$\alpha \equiv 2~(mod~4)$. By Theorem~\ref{thm:cond_fold2D},
$(\Lambda,\cS,\delta)$, $\delta=(+1,0)$, defines a folding. Now,
Theorem~\ref{thm:trans_shape} is used iteratively to form an
infinite $\cS'$-DCC, where $\cS'$ is "almost" a regular hexagon
with $\sqrt{a \cdot b} R + o(R)$ dots (the six vertices of one
hexagon are at $(\frac{\beta}{3},0)$, $(\beta,0)$,
$(\frac{4\beta}{3},\frac{\alpha}{2})$, $(\beta,\alpha)$,
$(\frac{\beta}{3},\alpha)$, $(0,\frac{\alpha}{2})$). Hence, a
lower bound on the number of dots in a regular hexagon with radius
$R$ is approximately $\frac{\sqrt{3\sqrt{3}}}{\sqrt{2}} R+o(R)$.

\subsection{Circle in the Square Grid}

We apply Theorem~\ref{thm:infinite}, where $\cS$ is a regular
hexagon with radius $\rho$ and $\cS'$ is a circle with radius $R$.
A lower bound on the number of dots in $\cS$ is approximately
$\frac{\sqrt{3\sqrt{3}}}{\sqrt{2}} \rho +o(\rho)$. The maximum on
$\frac{\sqrt{3\sqrt{3}} \rho+o(\rho)}{\sqrt{2}|\cS|} \Delta
(\cS,\cS')$ yields a lower bound of $1.70813R + o(R)$ on the
number of dots in $\cS'$.

\subsection{Other Shapes}

For most of the regular $n$-gons ($n \notin \{ 4,6 \}$) in the
square grid we applied Theorem~\ref{thm:mediator} with a hexagon
and a circle (for $n=4$ the optimum can be obtained from a Costas
array). Exceptions are $n=3,~5,~8,~10$, and 12, where we got a
better bound by specific constructions.
Table~\ref{tab:boundsummary} summarize the bounds we obtained for
regular polygons and a circle in the square grid. We also consider
circle in the hexagonal grid, but the main result in the hexagonal
grid is a construction of an optimal DDC whose shape is a hexagon.
This involves another interesting construction for a lattice
tiling of another shape in the square grid and translation into
the hexagonal grid. The same techniques can be used for any
$D$-dimensional shape. Finally, we note that the problem is of
interest also from discrete geometry point of view. Some similar
questions can be found in~\cite{LeTh95}.

\begin{table}
\caption{Bounds on the number of dots in an $n$-gon
DDC}\label{tab:boundsummary}
\begin{equation*}
\begin{array}{lccc}
\hline
$n$ & \text{upper bound} & \text{lower bound} & \text{ratio between bounds} \\
\hline
3 & 1.13975R & 1.02462R & 0.899   \\
4 &1.41421R & 1.41421R &  1  \\
5 & 1.54196R & 1.45992R  & 0.9468  \\
6 &1.61185R & \approx 1.61185R & \approx 1  \\
7 & 1.65421R & 1.55233R & 0.9384  \\
8 &1.68179R & 1.60094R & 0.9519 \\
9 & 1.70075R & 1.61343R & 0.9487  \\
10 &1.71433R & 1.64302R & 0.9584  \\
11 &1.72439R & 1.64458R & 0.9537  \\
12 &1.73205R & 1.66871R & 0.9634  \\
13 &1.73802R & 1.66257R & 0.9566  \\
14 &1.74275R & 1.66883R & 0.9576  \\
30 &1.76598R & 1.69955R & 0.9593  \\
90 &1.77173R & 1.70718R & 0.9636  \\
\text{circle} &1.77245R & 1.70813R & 0.9637 \\
\hline
\end{array}
\end{equation*}
\end{table}

\section{Application in Error-Correction}
\label{sec:ECC}

Assume that we have a $D$-dimensional array of size $n_1 \times
n_2 \times \cdots \times n_D$ and we wish to correct any
$D$-dimensional burst of length 2 (at most two adjacent positions
are in error). The following construction given in~\cite{YaEt09}
is based on folding the elements of a Galois field with
characteristic 2 in a parity check matrix, where the order of the
elements of the field is determined by a primitive element of the
field.

\noindent {\bf Construction A:} Let $\alpha$ be a primitive
element in GF($2^m$) for $2^m-1 \geq \prod_{\ell=1}^D n_\ell$. Let
$d=\lceil \log_2 D \rceil$ and $\bi=(i_1,i_2,\ldots,i_D)$, where
$0 \leq i_\ell \leq n_\ell-1$. Let $A$ be a $d\times D$ matrix
containing distinct binary $d$-tuples as columns. We construct the
following $n_1 \times n_2 \times \cdots \times n_D \times (m+d+1)$
parity check matrix $H$.

$$h_{\bi}=
\left[\begin{array}{c}
1 \\
A\bi^T \mod 2 \\
\alpha^{\sum_{j=1}^D i_j (\prod_{\ell=j+1}^D n_\ell)}
\end{array}\right].
$$
for all $\bi=(i_1,i_2,\ldots,i_D)$, where $0 \leq i_\ell \leq
n_\ell-1$.

\begin{theorem}
The code constructed in Construction A can correct any 2-burst in
an $n_1 \times n_2 \times \cdots \times n_D$ array codeword.
\end{theorem}
\begin{theorem}
The code constructed by Construction A has redundancy which is
greater by at most one from the trivial lower bound on the
redundancy.
\end{theorem}

The same construction will work if instead of a $D$-dimensional
array our codewords will have have a shape $\cS$ of size $2^m-1$
and there is a lattice tiling $\Lambda$ and a direction $\delta$
such that $(\Lambda,\cS,\delta)$ defines a folding. The elements
of GF($2^m$) will be ordered along the folded-row of $\cS$. The
construction can be generalized for more complicated types of
multidimensional errors.

\section{Application in Pseudo-Random Arrays}
\label{sec:pseudo-random}

Let $n=2^{k_1k_2}-1$ such that $n_1=2^{k_1}-1$ and $n_2 =
\frac{n}{n_1}$ are relatively primes and greater than 1. Let
$S=s_0s_1 \cdots s_{n-1}$ be an m-sequence (maximal length linear
shift register sequence~\cite{McSl76,Golomb}) of length $n$.
Assume we use F1 to form an $n_1 \times n_2$ array $\cA$. $\cA$
has many interesting properties such as shift, recurrences,
addition, auto-correlation, etc.~\cite{McSl76}. It also has a $k_1
\times k_2$ window property, i.e., each $k_1 \times k_2$ possible
binary matrix appears exactly once as a window in the cyclic
array. These arrays were called in~\cite{McSl76} pseudo-random
arrays. All these properties except for the window property are a
consequence of the fact that the elements in the folded-row are
consecutive elements of the m-sequence $S$. Hence, if we use any
of the folding operations to fold $S$ into a $D$-dimensional shape
$\cS$, the shape $\cS$ will have all these properties. As a
consequence of Theorem~\ref{thm:trans_shape} we have the following
theorem.
\begin{theorem}
\label{thm:window} Assume $\Lambda$ define a lattice tiling for an
$n_1 \times n_2$ array, such that $n_1 n_2 = 2^{k_1k_2}-1$, $n_1$,
$n_2$ are relatively primes and greater than 1. Assume further
that $\Lambda$ defines a lattice tiling for the shape $\cS$ and
$(\Lambda,\cS,\delta)$ defines a folding for a direction $\delta$.
Then, if we fold an m-sequence $S$ into $\cS$ by the direction
$\delta$ then the resulting shape $\cS$ has the $k_1 \times k_2$
window property if and only if the $n_1 \times n_2$ array $\cA$
has the $k_1 \times k_2$ window property by folding $S$ into $\cA$
by the direction $\delta$.
\end{theorem}

\section{conclusion}
\label{sec:conclude}

The well-known definition of folding was generalized. The
generalization makes use of a lattice tiling and a direction in
which the folding is performed. We demonstrated how folding in
general and the new definition in particular is applied for
constructions of multidimensional synchronization patterns,
error-correcting codes, and pseudo-random arrays. The compressed
discussion we have made raised lot of problems for further
research, which can advance the research on multidimensional
coding. It also raised intriguing questions which are related to
discrete geometry.

\section*{Acknowledgment}
This work was supported in part by the United States-Israel
Binational Science Foundation (BSF), Jerusalem, Israel, under
Grant No. 2006097.

%\bibliography{allbib,extra}

\end{document}